\documentclass[prb,preprint,showpacs,preprintnumbers,amsmath,amssymb]{revtex4}

\usepackage{graphicx,color}
\usepackage{dcolumn}
\usepackage{bm}

\begin{document}

\title{The influence of a single defect in composite gate insulators on the performance of nanotube transistors}

\author{YU Wen-Juan and WANG Neng-Ping\footnote{E-mail: wangnengping@nbu.edu.cn}}
\affiliation{Science Faculty, Ningbo University, Fenghua Road 818, 315211 Ningbo, P.R. China}

\date{\today}
\begin{abstract}
\textit{ The current through a carbon nanotube field-effect transistor (CNFET) with cylindrical gate 
electrode is calculated using the nonequilibrium Greens function method in a tight-binding approximation.
The obtained result is in good agreement with the experimental data.
The space radiation and nuclear radiation are known to cause defects in solids.
The theoretical approach is used to calculate the amplitude of the random-telegraph-signal (RTS) noise
due to a single defect in the gate oxide of a long channel p-type CNFET.
We investigate how the amplitude of the RTS noise is affected by the composite structure of 
gate insulators, which contains an inner insulator with a large dielectric constant ($\epsilon > 3.9$) and
an outer insulator with a dielectric constant of 3.9 (as for SiO$_2$).
It is found that the RTS amplitude increases apparently 
with the decreasing thickness of the inner gate insulator.
If the inner insulator is too thin, even though its dielectric constant is as large as 80, the amplitude 
of the RTS noise caused by the charge of Q = +1e may amount to around 80\% in the turn-on region.
Due to strong effects of defects in CNFETs, CNFETs have a potential to be used for detecting 
the space radiation or nuclear radiation.
}\\
\\
Key\ words:\ {carbon nanotube, defects, field-effect transistor, nonequilibrium Greens function, radiation effect
}
\end{abstract}


\maketitle

\section{Introduction}
\label{sec_1}

Carbon nanotubes (CNTs) have exceptional strength and stability, and they can exhibit either
metallic or semiconducting properties depending on the rolling direction (chirality)
and diameter[\onlinecite{Mintmire}].
Due to these structural and electronic properties, many studies have expected the applications of CNTs 
in future nanoelectronic devices[\onlinecite{Zhang,Wind,Javey03}].
Attention has focused especially on field-effect transistors (FETs).
With the continuous improvement of scaling and contacts,
high performances of carbon nanotube field-effect transistors (CNFETs) 
have been reported[\onlinecite{Wind,Javey03,Derycke,Chen,Franklin}].

CNFET has a significant potential to replace conventional silicon-based circuitry due to its
smaller size,  better electrostatics, and higher
mobility. And nanoelectronic devices will promisingly be used in future space technology and weapons technology.
The space radiation and nuclear radiation are known to cause defects in solids[\onlinecite{Dienes,Dartyge,
Gusarov,Luneville,Jin,Narayanan}]. 
And the fabrication process 
of CNFETs introduces also defects in the oxide or at the nanotube-oxide interface[\onlinecite{Chan}].
A defect in gate oxide may trap a current carrier on a carbon nanotube and then detrap it.
A charge state of the defect may keep for some time and then change into another charge state due to the
random trapping or detrapping. Such random charge-state switching
will cause a random current switching between two (or more) levels, i.e., random telegraph signals
(RTSs)[\onlinecite{Rall,Kirton,Uren,Liu06}].

RTSs due to individual defects in CNFETs have been observed in experiments[\onlinecite{Liu,Liu06,Liu08,Chan,Lee}],
and giant RTS values of 60\% or more have recently been detected[\onlinecite{Liu08}]. 
Some experiments[\onlinecite{Chan}] measured RTSs even at low drain bias of 50 mV due to defects
at the nanotube-oxide interface or in the oxide.
Theoretical studies have also indicated that a single defect in the gate dielectric of a CNFET may cause
a substantial current change in CNFETs[\onlinecite{Heinze,Neophytou,Wang}].
When there are many defects with different time constants, the superposition of RTS leads to
a low frequency 1/f noise[\onlinecite{Dutta, Weissman,Collins,Lin,Ishigami}].
Therefore,  CNFETs need not only to be provided with the resistance to radiations but also to be designed with low noise.
Clearly, studying the effect of individual defects is important for understanding the microscopic aspect of
1/f noise in CNFETs  as well as for designing a CNFET with low noise.

SiO$_2$ is a commonly used gate dielectric for the conventional metal-oxide-semiconductor field-effect transistors 
(MOSFETs) and CNFETs. To improve continuously the performances of the FETs,
the thickness limit of SiO$_2$ will soon be reached and a new dielectric material with a 
high dielectric constant is needed to replace SiO$_2$. The FET gate stack scaling with various
materials of high dielectric constant have widely been studied[\onlinecite{Wilk,Gusev}].
Usually, the band gap decreases as the dielectric constant increases.
Simply employing thinner HfO$_2$ ($\epsilon \approx$ 16) or other materials
with high dielectric constant is not a solution because the thinner HfO$_2$ or 
other materials of high dielectric constant will cause unacceptably high levels of
gate leakage current[\onlinecite{Lee}].

TiO$_2$ has a large dielectric constant that ranges from 40 to 86[\onlinecite{Brown,Rausch,Fuyuki}] and
its band gap is between 3.0 $\sim$ 3.5 eV[\onlinecite{Pascual,Fuyuki}],
depending on the crystalline structure. A considerable gate leakage current occurs if TiO$_2$-only 
gate dielectric is used due to small band gap and thermionic emission.
SiO$_2$ has a band gap of 9 eV, which is much larger than the band gap of TiO$_2$.
It has been known that composite gate oxides with TiO$_2$ and SiO$_2$ can reduce effectively the gate leakage
current within a range of appropriate gate voltages[\onlinecite{Campbell,Kim}].
Recently, composite gate dielectrics have been studied to reduce gate leakage current[\onlinecite{Campbell,Lu,Cho,Choi}]. 

As pointed out in Ref.[\onlinecite{Wang}], using a sigle gate oxide 
with large dielectric constant or small thickness
can reduce the RTS noise amplitude. However, for a compiste structure of gate insulators how the RTS noise 
amplitude depends on dielectric constants and thicknesses of inner and outer insulators is unknown up to now.
In this work, we will investigate systematically how the composite gate dielectrics affect
the amplitude of the RTS noise. This paper is organized as follows.
In section \ref{sec_2} we show the device setup for simulations
and give a brief overview of our methodology.
In section \ref{sec_3} we present our calculation results 
for the dependence of the RTS noise amplitude on
the nanotube radius and parameters of composite gate dielectrics,
and discuss also how to reduce the amplitude of the noise caused by  a charge in composite gate dielectrics.
Finally, conclusions will be given in section \ref{sec_4}.

\section{Methodology}
\label{sec_2}

The insets of Fig.\ref{current-exp} show the cylindrical geometry of a CNFET considered for calculations,
which is similar to the device used in Refs.[\onlinecite{Leonard,Heinze,Wang}].
In the present work, we consider a zigzag (n,0) semi-conducting carbon nanotube, 
where the index n will be defined in the calculation.
Source and drain leads are separated from the CNT in the radial
direction by a van der Waals distance of 0.3 nm.
The CNT is surrounded by a dielectric oxide, whose dielectric constant and thickness will be defined according
to the request of calculations.
The CNT is divided into three regions: semi-infinite source and drain regions having uniform potentials,
and a self-consistent scattering region.
The scattering lengths of the CNT inside the source and drain leads
are both defined as 24 nm. The channel length
between the source and drain leads is 200 nm unless otherwise specified.
The thickness of the gate oxide will be defined below.
The work function of the metal source and drain leads is chosen to be 1 eV larger than 
the work function of the CNT, which is measured from midgap.
This gives an ohmic p-type contact,
as obtained in recent experiments[\onlinecite{Javey03,Appenzeller}].
The temperature is fixed at $290 K$.

At an appropriate gate voltage, if a drain bias is applied a source-drain current will occur.
The drain voltage is related to the Fermi levels of the source and drain leads by 
the expression of $V_{D} = (\mu_{S} - \mu_{D})/e$.
The electron (or hole) transport properties can be calculated using the non-equilibrium Greens function method.
The retarded Greens function of the scattering region plays an important role in the theory.
It can be expressed as[\onlinecite{Datta}]
\begin{equation}
\displaystyle\bigg[\ (E\ + i0^+)I\ -\ H\ -\Sigma^{r}_{S}(E)\
             -\Sigma^{r}_{D}(E)\ \bigg]G^{r}(E)\ =\ I,
\label{Green-function}
\end{equation}
where $I$ is the identity matrix, $H$ is the Hamiltonian for the scattering region of the CNT, and
$\Sigma^{r}_{S,D}$ denote retarded self-energies of the source and drain leads, respectively.
We describe the Hamiltonian of a CNT using a tight-binding approximation with one $\pi $
orbital per carbon atom[\onlinecite{Heinze,Wang,Leonard,Guo}].
The nonequilibrium Greens function can easily be calculated using a mode-based method in Ref.[\onlinecite{Guo}].

The charge distribution in the scattering region can be evaluated
using the density matrix[\onlinecite{Brandbyge,Rocha}].
Then, the discrete charges on the CNT are broadened by a Gaussian along the nanotube
direction and in the radial direction with $\sigma\ =\ $0.07 nm, which is about half the C-C bond length[\onlinecite{Heinze}].
The charge of the defect is approximately treated as a point charge, 
smeared by a Gaussian along the nanotube axis and in the
radial direction with $\sigma\ =\ $0.05 nm[\onlinecite{Heinze}], and is placed midway between the source and drain leads.
Scattering from the Coulomb potential of the charged defect is treated neglecting band mixing.

The potential along a CNT can be obtained from the solution of the Poisson equation
at a finite drain voltage[\onlinecite{Brandbyge}]. It is found from Eq.(\ref{Green-function}) that
the Greens function $G^{r}(E)$ will be changed due to the potential,
which is included in the Hamiltonian $H$ of the CNT.
Therefore, the Greens function of the scattering region $G^{r}(E)$,
the charge on the CNT, and the potential along the CNT
must be determined self-consistently.

If the Greens function $G^{r}(E)$ is obtained from self-consistent calculations,
one can calculate the current through the system using the well-known
Landauer-B\"{u}ttiker formula[\onlinecite{Buettiker,Meir,Datta}]
\begin{equation}
I\ =\ \displaystyle\frac{2e}{h}\int dE T(E)
                   \displaystyle\bigg[\ n_{F}(E - \mu_{S})\
                    -\ n_{F}(E - \mu_{D}) \ \bigg]\ ,
\label{current}
\end{equation}
where the factor of 2 is due to the degeneracy of spin and $n_{F}$ is the Fermi function.
The transmission between the source and drain leads in Eq.(\ref{current}) 
can be determined using the following
expression[\onlinecite{Meir,Datta,Svizhenko}]
\begin{equation}
T(E)\ =\ tr\displaystyle\bigg( \Gamma_{S}(E)G^{r}(E)
                              \Gamma_{D}(E)G^{a}(E)
                              \displaystyle\bigg)\ ,
\label{transmission}
\end{equation}
where $\Gamma_{S,D}$ is defined by the formula of 
$\Gamma_{S,D}\ =\ i
  \displaystyle\bigg[\ \Sigma^{r}_{S,D}(E)\ -\ \Sigma^{a}_{S,D}(E)\ \bigg] $.

For a p-type CNFET, the hole emission and capture by a defect in the gate oxide
will cause randomly the current switchings between $I_0$ and $I_Q$, i.e. RTSs.
The amplitude of such RTS noise is defined as the relative 
current change due to the charge of the defect[\onlinecite{Wang}]
\begin{eqnarray}
A_{RTS}=(I_0-I_Q)/I_0\ ,
\end{eqnarray}
where $I_Q$ and $I_0$ are the current with and without the charge.

\section{Numerical Results}
\label{sec_3}

In this section, we investigate in detail how the amplitude of the RTS noise due to a single defect 
depends on the nanotube radius and how the RTS noise amplitude relys on the parameters of composite gate dielectrics, 
such as, thicknesses and dielectric constants of the inner insulator and the outer insulator.
The nanotube-direction position of the charge is fixed in the middle of the channel.

\subsection{\bf Dependence of $A_{RTS}$ on $R_{CNT}$}
\label{sec_3a}

\begin{figure}
\vspace{0.3cm}
\begin{center}
\includegraphics[scale=0.8,angle=0]{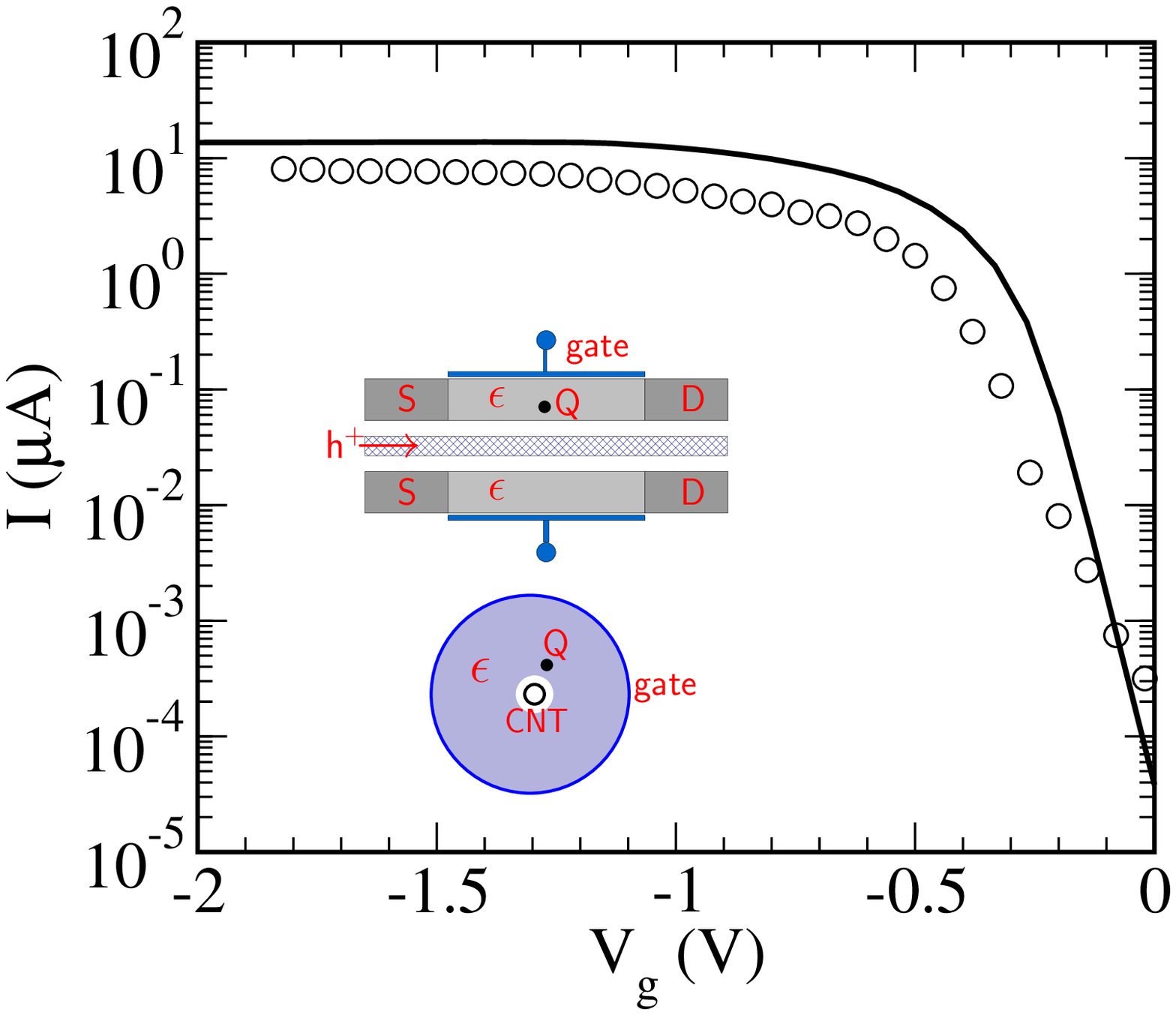}
\caption{
The calculated current (solid curve) at the drain voltage of -0.1 V is compared with the experimental
result (open circles)[\onlinecite{Franklin}].
The calculation is done for a CNFET with a zigzag (14,0) CNT
($R_{CNT}\ =\ 0.56$ nm) and a 10 nm-thick HfO$_2$ gate dielectric. The channel length between source and
drain electrodes is defined as 40 nm.
Insets show cross-sectional plots for the cylindrical geometry of the device used in
the calculation.
}
\label{current-exp}
\end{center}
\vspace{0.3cm}
\end{figure}

To demonstrate the validity of our model, we calculate the current through a
CNFET of sylindrical geometry with a zigzag (14,0) carbon nanotube and a 10 nm-thick HfO$_2$ dielectric.
The channel length in the calculation is defined as 40 nm. The obtained result for $V_d = -0.1$ V is shown
in Fig.\ref{current-exp}. A recent experiment[\onlinecite{Franklin}] measured the current at $V_d = -0.1$ V through
a CNFET with a cylindrical geometry of gate electrode. In the experiment, the carbon nanotube was 1 nm in diameter
and 40 nm in length, and the gate oxide was 10 nm-thick HfO$_2$.
The experimental result is also displayed in Fig.\ref{current-exp} for comparison.
Clearly, our calculation results are in quite good agreement with experimental
data[\onlinecite{Franklin}]. 
The good agreement between theoretical results and experimental data justifies our model.

\begin{figure}
\vspace{0.3cm}
\begin{center}
\includegraphics[scale=0.6,angle=0]{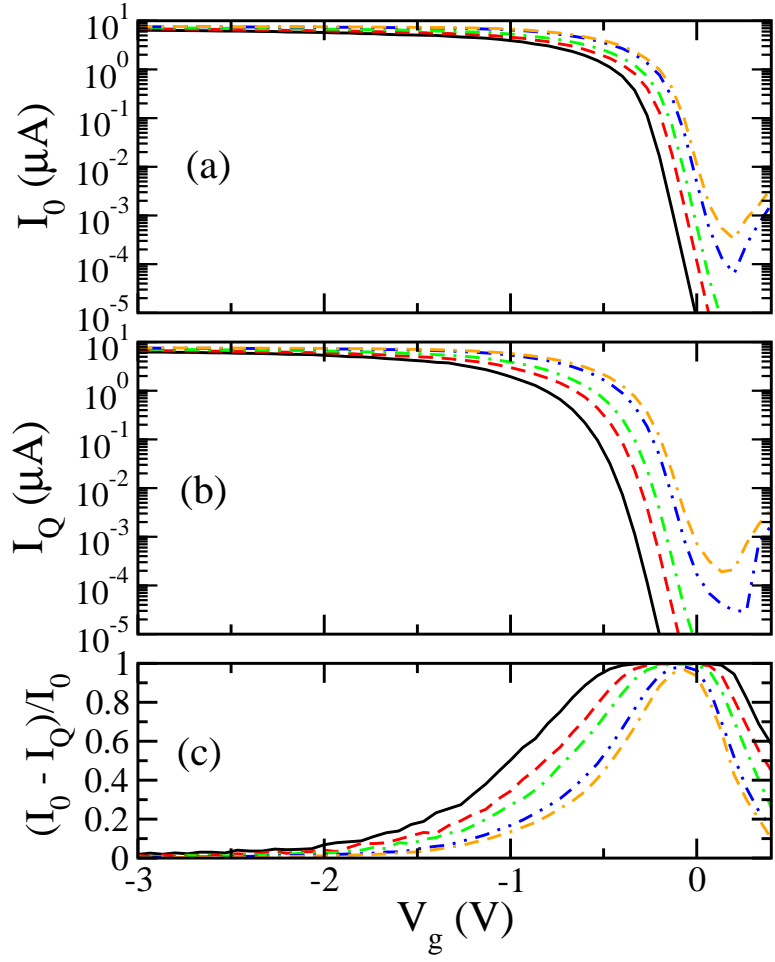}
\vspace{-0.2cm}
\caption{(Color online)
Effect of trapped charge on current through a transistor for
different zigzag (n,0) carbon nanotubes with the index n being 13 (black solid curve), 17 (red dashed curve),
20 (green dot-dashed curve), 25 (blue double dot-dashed curve),
and 29 (orange double dash-dotted curve). The calculations are
performed for drain voltage of -50 mV, gate radius of 16 nm,
and gate oxide of SiO$_2$. (a) Current vs gate voltage with no trapped charge. 
(b) Same as (a) for a positive charge
of Q=+1e at the nanotube-oxide interface.
(c) The amplitude of the RTS noise due to the charge of Q=+1e at the
radial position defined in (b).}
\label{diff_nanotube}
\end{center}
\vspace{0.3cm}
\end{figure}

We consider a commonly used insulator of SiO$_2$ ($\epsilon = 3.9$) as a gate oxide of a CNFET
and investigate the dependence of
$A_{RTS}$ on the nanotube radius.
For different zigzag (n,0) carbon nanotubes, where the index n is 13, 17, 20, 25, and 29,
we calculate the current through the device at the drain voltage of -50 mV and at different gate voltages from -3 V to 0.5 V. 
For the gate radius of $R_g = 16$ nm, the obtained result is shown 
in panel (a) of Fig.\ref{diff_nanotube} as a function of the applied gate voltage.
The (13,0) zigzag nanotube gives the large value of on-off current ratio, $I_{on}/I_{off}\ \approx 10^{11}$, while
the (29,0) zigzag nanotube gives a much lower on-off current ratio, $I_{on}/I_{off}\ \approx 10^4$.

For a positive charge of Q = +1e located at a radial distance of 0.4 nm from the nanotube wall (i.e., at the nanotube-oxide 
interface), we calculate the amplitude of the RTS noise due to the positive charge for a gate radius of $R_g\ =\ 16$ nm,
a dielectric constant of 3.9, and zigzag (n,0) carbon nanotubes with the index n being 13, 17, 20, 25, and 29.
From the calculated current through a CNFET with and without charged defect in the device,
we calculate the amplitude of the RTS noise due to the charged defect. 
The obtained results are shown as a function of gate voltage $V_g$ in Fig.\ref{diff_nanotube}.
It is clear that the smaller the nanotube is, the bigger the amplitude of the RTS noise is.

\begin{figure}
\vspace{0.3cm}
\begin{center}
\includegraphics[scale=0.6,angle=0]{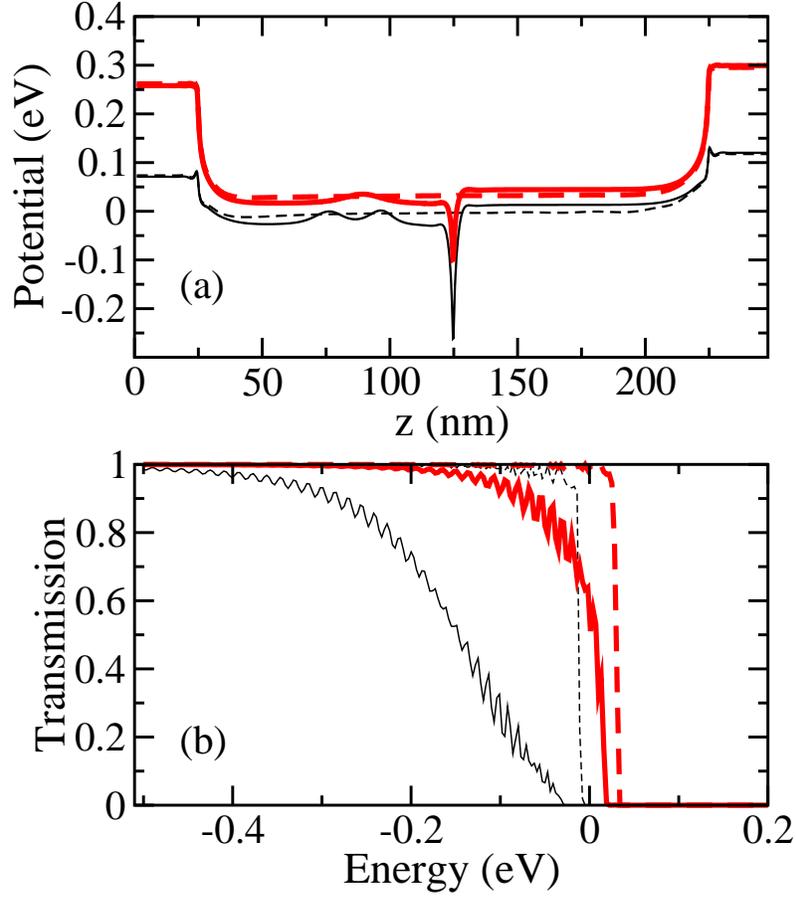}
\vspace{-0.2cm}
\caption{(Color online)
A comparison of the potential (or transmission) for the device considered
in Fig.\ref{diff_nanotube} with a zigzag (13,0) CNT and for that with a zigzag (29,0) CNT
at a gate voltage of V$_g$ = -0.5 V and a drain voltage of V$_d$ = -50 mV.
(a) Potential along the (13,0) nanotube for Q = 0
(thin black dashed curve) and for Q=+1e (thin black solid curve) at the nanotube-SiO$_2$ interface
and potential along the (29,0) nanotube for Q = 0
(thick red dashed curve) and for Q=+1e (thick red solid curve) at the nanotube-oxide interface.
(b) Transmission for those four cases defined in (a). }
\label{poten-trans}
\end{center}
\vspace{0.3cm}
\end{figure}

In order to understand the dependence of the RTS noise amplitude on the nanotube radius, 
we calculate the potential along the nanotube and show the result
in Fig.\ref{poten-trans}(a).
In the absence of any scatterer, the potential in the channel is almost constant in midchannel, as expected for
a long-channel ballistic device. A positively charged trap right at the nanotube-oxide interface causes an extra barrier
for hole transport, giving a shift in the plot of transmission vs energy in Fig.\ref{poten-trans}(b).
This shifts the voltage threshold of the device, as seen in Fig.\ref{diff_nanotube}.
Comparing the potential for the (13,0) nanotube and the potential for the (29,0) nanotube,
concerning the hole transmission the potential barrier of the former
is much larger than that of the latter.
This gives a less abrupt turn-on of transmission with energy
for the (13,0) nanotube as shown in Fig.\ref{poten-trans}(b).
Therefore, we expect a larger amplitude of the RTS noise for the (13,0) nanotube.

From the above calculations of current through idea CNFETs (i.e., without any charged defect) for different
carbon nanotube radii, one can see that
a smaller semi-conducting nanotube gives a larger value of on-off current ratio $I_{on}/I_{off}$.
However, the RTS noise amplitude for a smaller semi-conducting nanotube is larger, 
which suggests one to make an intermediate choice
between different nanotube radii.
In the following, we use a zigzag (17,0) carbon nanotube, which has a radius of 0.67 nm, to study the RTS noise in FETs.

\subsection{\bf Role of composite gate insulators}
\label{sec_3b}

The amplitude of the RTS noise is known to decrease with the gate dielectric thickness. 
However, if the gate dielectric is too thin the gate leakage current will occur because of direct tunneling
between the bands of the semi-conducting nanotube and 
the Fermi level of the gate electrode. 
Composite gate insulators may be used to prevent or reduce 
the gate leakage current[\onlinecite{Campbell,Kim}].
The question then arises: how does the amplitude of the RTS noise due to a charged defect 
depend on the parameters of the composite gate insulators?
To answer this question, we will investigate the current reduction due to a charged defect in
a composite gate insulators as
shown in Fig.\ref{device_setup_2}. For convenience, we denote the thickness and dielectric constant of the inner insulator
as $t_1$ and $\epsilon_1$, respectively. The outer insulator is supposed as SiO$_2$, 
whose thickness and dielectric constant are denoted by $t_2$ 
and $\epsilon_2$ ($\epsilon_2$ = 3.9), respectively.

As an example, we use 1 nm-thick HfO$_2$
for the inner insulator (i.e., $\epsilon_1 = 16, t_1 = 1$ nm) and 2 nm-thick SiO$_2$ for the outer
insulator (i.e., $\epsilon_2 = 3.9, t_2 = 2$ nm) in the composite structure. We calculate the current through the device
without any charged defect (Q = 0) and the current for a charge of Q = +1e located at 
the nanotube-insulator interface. 
From the results of the current $I_0$ for Q = 0 and the current $I_Q$ for
Q = +1e,  we calculate the relative current reduction $(I_0 - I_Q)/I_0$ due to the charge of Q = +1e.
The obtained result of the current $I_0$ is shown as black solid curve in panel (a) of 
Fig.\ref{current_cgi_eps16}. The calculated relative current reduction is shown as black solid curve in panel (b) of 
Fig.\ref{current_cgi_eps16}. The relative current reduction due to the charge of Q = +1e 
is estimated as 91\% in the turn-region and 4\% in the on-state. 
We enlarge the thickness of the inner gate insulator HfO$_2$ to 13 nm (i.e., $\epsilon_1 = 16, t_1 = 13$ nm)
and keep the same thickness of the outer insulator SiO$_2$ 
as before (i.e., $\epsilon_2 = 3.9, t_2 = 2$ nm). The current for Q = 0 and Q = +1e 
and the relative current reduction due to the charge of Q = +1e are calculated. 
The obtained result of the current for Q = 0 and
that for the relative current reduction are shown in  panels (c) and (d) of 
Fig.\ref{current_cgi_eps16}, respectively. 
The relative current reduction in the turn-on region is around 80\%,
which is decreased by about 11\% as compared with the corresponding result for $t_1 = 1$ nm obtained before.
The comparison of panels (b) and (d) of Fig.\ref{current_cgi_eps16} represents that the relative current reduction
due to a charge in composite gate dielectrics increases with the decreasing total thickness of composite gate dielectrics
if the outer insulator is unchanged.
For comparison sake, the results of relative current reduction for 3 nm-thick SiO$_2$ and 
3 nm-thick HfO$_2$ are shown in panel (b) of Fig.\ref{current_cgi_eps16}, 
and the corresponding results for 15 nm-thick SiO$_2$ and 15 nm-thick HfO$_2$ are shown in panel (d) of Fig.\ref{current_cgi_eps16}.
As expected, the relative current reduction due to a charge in single gate dielectric of SiO$_2$ or HfO$_2$ 
decreases if the gate dielectric gets thinner.
Apparently, the dependence of the relative current reduction due to a charge in composite gate dielectrics 
on the thickness of the composite gate dielectrics is contradictory to the
dependence of the relative current reduction caused by a charge in a single gate dielectric 
on the thickness of the single gate dielectric.

To see more clearly, we consider another example of composite gate dielectrics, which
is a combination of the inner insulator with $\epsilon_1 = 40, t_1 = 1 $ nm and 
the outer insulator of 2 nm-thick SiO$_2$ (namely $\epsilon_2 = 3.9, t_2 = 2 $ nm).
We calculate the current through the device
for Q = 0 and Q = +1e. From the results of the current $I_0$ for Q = 0 and the current $I_Q$ for 
Q = +1e,  we calculate the relative current reduction due to the charge of Q = +1e.
The obtained results of the current $I_0$ and 
the relative current reduction are shown in panels (a) and (b) of Fig.\ref{current_cgi_eps40}, respectively.
To examine the dependence of the relative current reduction on the thickness of the inner insulator, we
calculate also the current for composite insulators with $\epsilon_1 = 40, t_1 = 13 $ nm 
and $\epsilon_2 = 3.9, t_2 = 2 $ nm. 
The results for the current $I_0$ and the relative current reduction are shown 
in in panels (c) and (d) of Fig.\ref{current_cgi_eps40}, respectively.  
The relative current reduction $(I_0 - I_Q)/I_0$ for $t_1 = 1 $ nm is estimated as 84\% in the turn-on region
and 0.4\% in the on-state.
And the relative current reduction decreases apparently to 49\% in the turn-on region 
and 0.1\% in the on-state if $t_1 = 13 $ nm.

\begin{figure}
\vspace{0.3cm}
\begin{center}
\includegraphics[scale=0.5,angle=0]{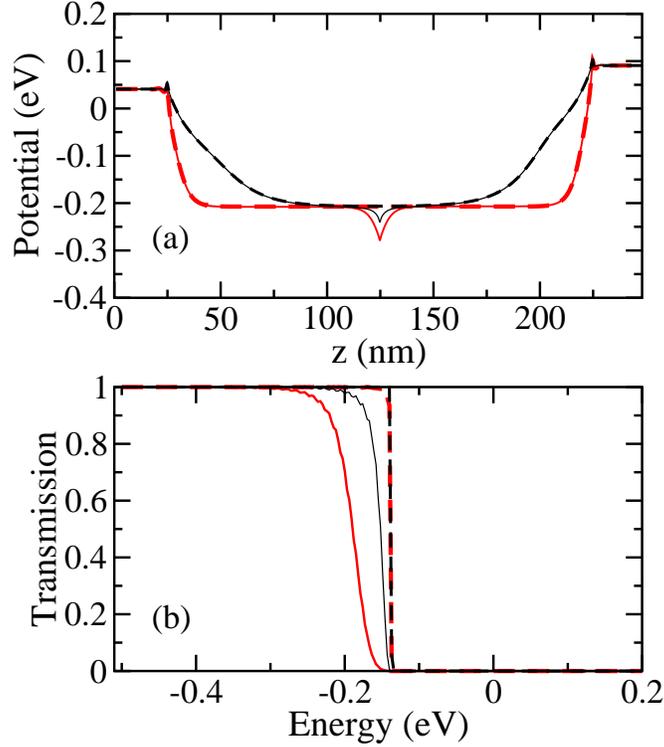}
\vspace{-0.2cm}
\caption{(Color online)
Panel (a) shows the potential along the (17,0) CNT for Q = 0 (red dashed curve) and for
Q = +1e (red soild curve) in the composite gate insulators, which contains the 1 nm-thick inner-dielectric
with a dielectric constant of 40 and the 2 nm-thick outer-dielectric SiO$_2$,
and the potential along the nanotube for Q = 0 (black dashed curve)
and for Q = +1e (black solid curve) in the composite gate insulators, which include the 13 nm-thick inner-dielectric
having a dielectric constant of 40 and the 2 nm-thick outer-dielectric SiO$_2$.
Panel (b) shows the transmission for those four cases defined in panel (a).
All results are obtained at a gate voltage of $V_g = -0.1$V and a drain volatge of 
$V_d = -50$ mV.}
\label{poten-trans_1}
\end{center}
\vspace{0.3cm}
\end{figure}

To understand why the relative current reduction due to a charge in composite gate dielectrics 
increases with the decreasing total thickness of composite gate dielectrics, 
in Fig.\ref{poten-trans_1} we show the potential along the nanotube for Q = 0  and for Q = +1e 
located at the nanotube-insulator interface of a CNFET with the composite gate insulators.
Clearly, as the inner insulator thickness $t_1$ increases from 1 nm to 13 nm, 
the potential barrier for hole transport due to Q = +1e decreases. 
The screening of the impurity charge contains the screening by the composite gate insulators
and the screening by the gate electrode. 
From the top panel of Fig.\ref{poten-trans_1}, it is found that 
the total screening effect of the impurity charge in the composite gate insulators
decreases if the full thickness of the composite gate dielectrics becomes smaller and
the thickness of the outer insulator maintains unchanged. 
The screening of the impurity charge by the gate electrode increases
with the decreasing thickness of the composite gate dielectrics because the screening by the gate electrode
becomes stronger when the gate electrode is closer to the impurity charge[\onlinecite{Wang}].
The screening by the composite gate dielectrics is dependent on the effective 
dielectric constant of the composite gate dielectrics. 

In order to know the screening of the impurity charge by the composite gate dielectrics,
we discuss the effective dielectric constant of the composite insulators.
We consider the device in Fig.\ref{device_setup_2} as a cylindrical capacitor, which contains 
a solid cylindrical conductor of radius $R_0 \approx R_{CNT} + 0.3 \approx 1$ nm, surrounded by a coaxial cylindrical conductive
shell of inner radius $R_2 = R_g$. The length of both cylinders is the channel length $L_c$ 
and we take this length to be much larger than $R_2 - R_0$. If there is no gate dielectric,
the capacitance of the device is expressed as $C_0$. And then, we suppose the two insulators with dielectric
constants of $\epsilon_1$ and $\epsilon_2$
($\epsilon_1 > \epsilon_2$) exist between the two cylindrical conductors and the radius of the interface between the
two insulators is defined as $R_1$. In this case, the corresponding capacitance is written as $C_x$. The effective
dielectric constant of the composite insulators may be defined as $\bar{\epsilon} = C_x/C_0$. If the length of both conductors
in the capacitor is much larger than the distance between the two conductors, i.e., $L_c >> R_2 - R_0$,
the effective dielectric constant may be estimated as
\begin{equation}
\begin{array}{rcl}
\bar{\epsilon} & \ = \ & \displaystyle \left[\frac{1}{\epsilon_1} + 
       ( -\frac{1}{\epsilon_1} + \frac{1}{\epsilon_2} )\ln\frac{R_2}{R_1}
       /\ln\frac{R_2}{R_0} \right]^{-1} \\ 
             \ & \ = \ & \displaystyle \left[\frac{1}{\epsilon_1} + 
       ( -\frac{1}{\epsilon_1} + \frac{1}{\epsilon_2} )\ln(1 + \frac{t_2}{R_0 + t_1})
       /\ln(1 + \frac{t_1}{R_0} + \frac{t_2}{R_0}) \right]^{-1} \ ,
\end{array}
\label{eff_dielec}
\end{equation}
where $\epsilon_1$ and $\epsilon_2$ are the dielectric constants of the inner and outer insulators, respectively,
and $t_1$ and $t_2$ are the thicknesses of the inner and outer insulators, respectively, i.e.,
$t_1 = R_1 - R_0$ and $t_2 = R_2 - R_1$.
One may use Eq.(\ref{eff_dielec}) to explain qualitatively the dependence of
the effective dielectric constant of composite gate insulators (as shown in Fig.\ref{device_setup_2}) on the
thickness of the inner insulator $t_1$.
It is clear from Eq.(\ref{eff_dielec}) that if $\epsilon_1 > \epsilon_2$, $t_1$ increases and $t_2$ keeps unchanged,
the effective dielectric constant of the composite insulators increases. 
So, the screening of the impurity charge by the composite insulators increases if the full thickness of the composite insulators
gets larger and the outer insulator maintains the same thickness. 
On the contrary, the screening of the impurity charge by the gate electrode decreases
with the increasing thickness of the composite insulators.
From results shown in the top panel of Fig.\ref{poten-trans_1}, we conclude that
the screening of the impurity charge by the composite insulators increases
with the full thickness of the composite insulators faster than the screening of the impurity charge by the gate electrode
decreases with the increasing full thickness of the composite insulators, 
so that the summing-up screening effect of the impurity charge increases
with the full thickness of the composite insulators 
if the thickness of the outer insulator is kept the same.

\begin{figure}
\vspace{-0.3cm}
\begin{center}
\includegraphics[scale=0.5,angle=0]{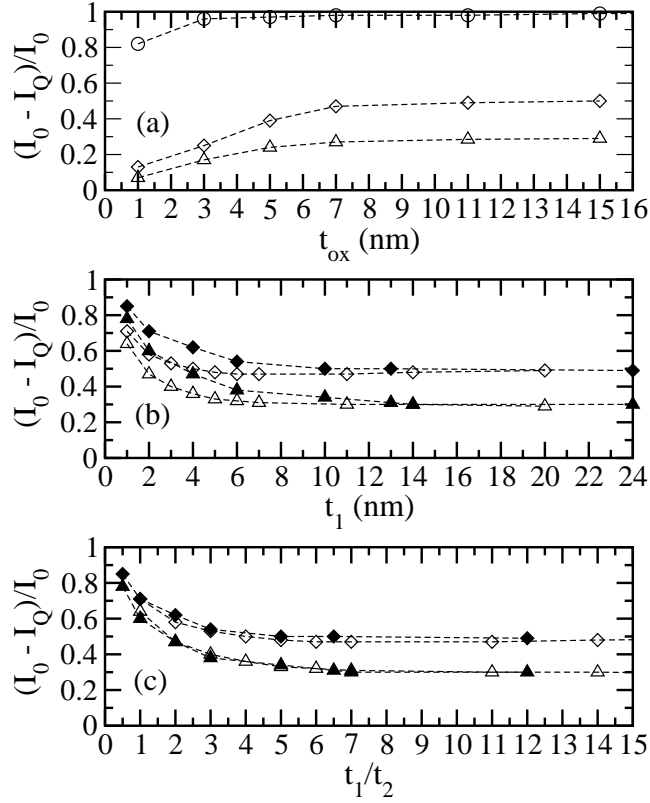}
\vspace{0.3cm}
\caption{The relative current reduction at the gate
overdrive of $V_g - V_{th,Q=0} = -0.1 V$
due to a charge of Q = +1e at a distance of 0.4 nm from the nanotube wall.
Panel (a) shows the relative current reduction as a function of gate oxide thickess
for different dielectric constants of
3.9 (open circles), 40 (open diamonds), and 80 (open triangles), respectively.
In panel (b), the relative current reduction for composite gate insulators, which contains
an inner insulator having a dielectric constant of 40 or 80  and an outer insulator of 1 nm or 2 nm-thick
SiO$_2$, is shown as a function of the inner insulator thickness $t_1$.
For the outer insulator of 1 nm-thick SiO$_2$,
the results obtained for the inner-insulator dielectric constant of
$\epsilon_1$ = 40  or 80 are shown as open diamonds or open triangles, respectively.
For the outer insulator of 2 nm-thick SiO$_2$,
the corresponding results for the inner-insulator dielectric constant of $\epsilon_1$ = 40 or 80  are shown as
solid diamonds or solid triangles, respectively.
Panel (c) shows the results for the composite insulators defined in panel (b),
as a function of the ratio of the inner insulator thickness $t_1$
to the outer insulator thickness $t_2$.
All the results are obtained for a drain voltage of -50 mV and a zigzag (17,0) CNT.
}
\label{RTS_cgi}
\end{center}
\vspace{0.3cm}
\end{figure}

Since the screening of the impurity charge in the composite insulators is sophisticatedly dependent on
the thickness of the inner insulator $t_1$ and the thickness of the outer insulator $t_2$,
we investigate how the relative current reduction depends on 
the thickness of the inner insulator $t_1$ within a wide range of $t_1$ 
when the thickness of the outer insulator is fixed.
We consider different thicknesses of the inner insulator from $t_1 = 1$ nm to $t_1 = 24$ nm and define
$\epsilon_2 = 3.9$ and $t_2 = 2$ nm.
And we calculate the relative current reduction at the gate voltage of $V_g = -0.1$ V, at which 
the relative current reduction has a maximal value as compared with 
the relative current reductions at other gate voltages (see panels (b) and (d) of Figs.\ref{current_cgi_eps16}
and \ref{current_cgi_eps40}).
In the middle panel of Fig.\ref{RTS_cgi}, we use solid diamonds to show the obtained results of
the relative current reduction $(I_0 - I_Q)/I_0$ as a function of the inner insulator thickness $t_1$.
For comparison, we show the relative current reductions at the
gate voltage of $V_g = -0.1$ V for a single gate insulator with a dielectric constant of 
3.9, 40, and 80 in the top panel of Fig.\ref{RTS_cgi} as a function of the thickness of the gate oxide.
And the relative current reduction for $\epsilon = 3.9, 40$, and $80$
increases nearly to the saturation value of 100\%, 49\%, and 30\%, respectively, 
if the thickness of the gate oxide $t_{ox} \stackrel{>}{\sim} 11$ nm.
From Fig.\ref{RTS_cgi}, one can find that for composite gate dielectrics 
the relative current reduction increases with the decreasing 
thickness of the inner insulator $t_1$. If $t_1 > 20$ nm, the relative current reduction
decreases approximately to the saturation value of 49\%, which is the same with the saturation value of
$(I_0 - I_Q)/I_0$ for a single gate oxide of $\epsilon = 40$. 
The reason is that if the thickness of the inner insulator $t_1$ is large enough,
the effective dielectric constant of the composite insulators $\bar{\epsilon}$ is nearly
the same with the $\epsilon_1$, as one can easily see from Eq.(\ref{eff_dielec}).
Furthermore, from the top panel of Fig.\ref{RTS_cgi} we can see that the relative current reduction for a single gate insulator
increases nearly to its saturation value if the thickness of the gate oxide $t_{ox} \stackrel{>}{\sim} 11$ nm, and this 
indicates that the screening of the impurity charge by the gate electrode becomes very weak if the thickness of the gate oxide
is large enough. 

Fixing the thickness of the outer insulator as $t_2\ =\ $ 1 nm, we calculate the relative current reduction 
for different thicknesses of the inner insulator $t_1$ and show 
the results of the relative current reduction at $V_g = -0.1$ V as a function of the inner insulator
thickness $t_1$ in the middle panel of Fig.\ref{RTS_cgi}.
To compare the results for the outer insulator thicknesses of $t_2 = 1$ nm and $t_2 = 2$ nm, 
in the bottom panel of Fig.\ref{RTS_cgi}
we show the results of the relative current reduction as a function of the ratio $t_1/t_2$. It is
found that for $t_2 = 1$ nm and $t_2 = 2$ nm, the dependences of the relative current reduction 
on the ratio $t_1/t_2$ are almost the same. 

For comparison sake, we consider a composite structure of a inner insulator with $\epsilon_1 = 80$
and an outer insulator with $\epsilon_2 = 3.9$ (as for SiO$_2$),
calculate the relative current reduction at $V_g = -0.1$ V for different inner-insulator thicknesses $t_1$
fixing the outer insulator thickness as $t_2 = 1$ nm (or $t_2 = 2$ nm), 
and the obtained results are shown as open triangles (or solid triangles) in Fig.\ref{RTS_cgi}.
It is found from Fig.\ref{RTS_cgi} that if $t_1/t_2 \stackrel{>}{\sim} 10$
the relative current reduction at $V_g = -0.1$ V reaches approximately saturation. And the saturation
value of the relative current reduction is 
around 49\% and 30\% for $\epsilon_1 = 40$ and $\epsilon_1 = 80$, respectively.
Such amplitudes of the RTS noise in the turn-on region are much smaller than 
the corresponding RTS noise amplitude for the SiO$_2$-only gate oxide.

\section{Conclusions}
\label{sec_4}

In this work, we calculate current through a ballistic p-type carbon nanotube transistor, using
the nonequilibrium Green function method in a tight-binding approximation.
The obtained current for an idea CNFET (i.e., without any charged defect) is in good agreement
with experimental result[\onlinecite{Franklin}].
From the calculated current through a CNFET with and without charged defect in the device,
we evaluate the amplitude of the random-telegraph-signal noise
due to the defect. 
We have investigated the dependence of the RTS noise amplitude due to a single postive charge
on the nanotube radius.
It is found that the RTS noise amplitude increases with the decreasing nanotube radius.

\begin{figure}
\vspace{-9.0cm}
\begin{center}
\includegraphics[scale=0.7,angle=0]{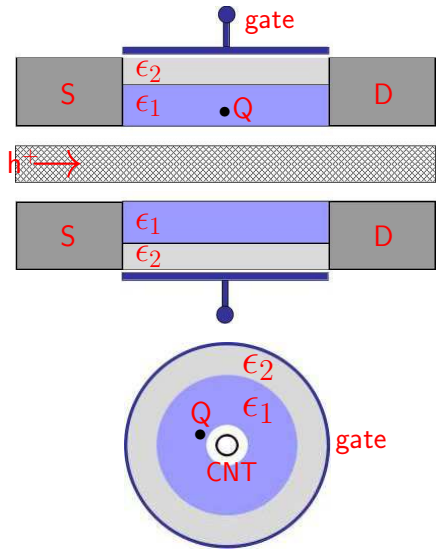}
\vspace{-8.0cm}
\caption{(Color online)
Cross-sectional plots for a CNFET with composite gate dielectrics.
The outer insulator is SiO$_2$ with a dielectric constant of $\epsilon_2 = 3.9$.
The dielectric constant of the inner insulator 
is defined as higher than that of SiO$_2$ ($\epsilon_1 > 3.9$).
}
\label{device_setup_2}
\end{center}
\vspace{0.0cm}
\end{figure}

\begin{figure}
\vspace{0.3cm}
\begin{center}
\includegraphics[scale=0.5,angle=0]{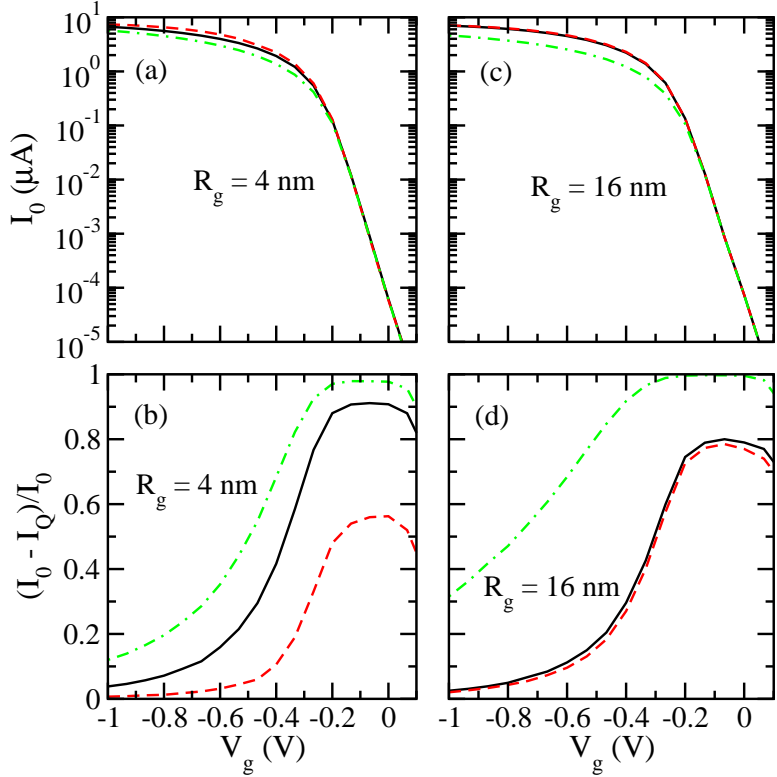}
\vspace{0.3cm}
\caption{(Color online) 
Panel (a) shows the current through an idea CNFET (i.e., without any impurity charge, Q =0) with a gate dielectric 
of 3 nm-thick SiO$_2$ (green dot-dashed curve), or with a gate dielectric of 3 nm-thick HfO$_2$ (red dashed curve),
or with a composite structure of gate insulators (black solid curve) as depicted
in Fig.\ref{device_setup_2}, in which the inner insulator is 1 nm-thick HfO$_2$ 
and the outer insulator is the 2 nm-thick SiO$_2$.
Panel (b) shows the relative current reductions 
due to a trapped charge of Q = +1e at the nanotube-insulator interface
for different gate dielectrics defined in panel (a).
Panel (c) shows the current through an idea CNFET (i.e., Q =0) for the same kinds of gate dielectrics as depicted in panel (a),
but the thickness of the single gate dielectric (SiO$_2$ or HfO$_2$) and that of the composite gate insulators are both 15 nm. 
For the composite gate insulators, the inner insulator is 13 nm-thick HfO$_2$
and the outer insulator is the 2 nm-thick SiO$_2$.
Panel (d) shows the relative current reductions 
due to a trapped charge of Q = +1e at the nanotube-insulator interface
for different gate dielectrics defined panel (c).
All the results are obtained at the drain voltage of -50 mV for a zigzag (17,0) CNT.
}
\label{current_cgi_eps16}
\end{center}
\vspace{0.3cm}
\end{figure}

\begin{figure}
\vspace{0.3cm}
\begin{center}
\includegraphics[scale=0.5,angle=0]{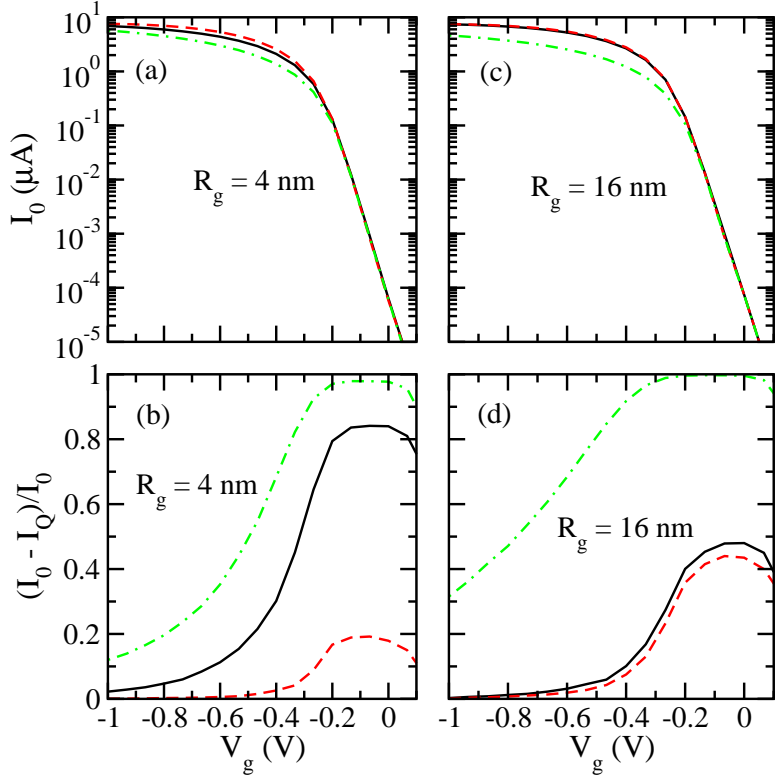}
\vspace{0.3cm}
\caption{(Color online) 
Panel (a) shows the current through an idea CNFET (i.e., without any impurity charge, Q =0) with a gate dielectric 
of 3 nm-thick SiO$_2$ (green dot-dashed curve), or with a gate dielectric having a thickness of 3 nm and a dielectric
constant of 40 (red dashed curve),
or with a composite structure of gate insulators (black solid curve) as depicted
in Fig.\ref{device_setup_2}, in which the inner insulator has a thickness of 1 nm and a dielectric constant of 40
and the outer insulator is the 2 nm-thick SiO$_2$.
Panel (b) shows the relative current reductions due to a trapped charge of Q = +1e at the nanotube-insulator interface
for different gate dielectrics defined in panel (a).
Panel (c) shows the current through an idea CNFET (i.e., Q =0)
for the same kinds of gate dielectrics as depicted in panel (a),
but the thickness of the single gate dielectric (SiO$_2$ or the dielectric having a dielectric constant of 40) 
and that of the composite gate insulators are both 15 nm.
For the composite gate insulators, the inner insulator is the 13 nm-thick dielectric with $\epsilon_1 = 40$
and the outer insulator is the 2 nm-thick SiO$_2$.
Panel (d) shows the relative current reductions due to a trapped charge of Q = +1e at the nanotube-insulator interface
for different gate dielectrics defined panel (c).
All the results are obtained at the drain voltage of -50 mV for a zigzag (17,0) CNT.
}
\label{current_cgi_eps40}
\end{center}
\vspace{0.3cm}
\end{figure}

We consider as gate dielectrics a composite structure of an inner insulator with a large dielectric constant
and an outer insulator with a dielectric constant of 3.9 (as for SiO$_2$) as shown in Fig\ref{device_setup_2}.
We have studied the relative current reduction due to a charged impurity in the composite gate insulators.
The screening of the impurity charge includes the screening of the impurity charge by the gate electrode
and the screening of the impurity charge by the composite gate insulators. 
The screening effect of the impurity charge by the gate electrode increases with the decreasing full thickness
of the composite gate insulators. On the contrary, the screening of the impurity charge by the composite gate insulators
diminishes if the full thickness of the composite gate insulators gets smaller and the thickness of the outer insulator remains the same.
This is because when the thickness of the outer insulator is fixed the effective constant constant of the composite gate insulators 
diminishes with the decreasing thickness of the inner insulator.
Since the former screening effect varies with the full thickness of the composite gate insulators slower than the latter 
screening effect, the summing-up screening effects of the impurity charge weakens with the decreasing full thickness
of the composite gate dielectrics if the thickness of the outer dielectric remains unaltered.

The relative current reduction due to the impurity charge in a p-type CNFET
increases with the decreasing ratio of the inner-insulator thickness to the outer-insulator thickness.
If the inner insulator is very thin, even though its dielectric constant is as large as 80, the amplitude
of the RTS noise caused by the charge of Q = +1e may amount to around 80\% in the turn-on region.
If the thickness of the inner gate insulator is much larger than the thickness of the outer gate insulator, 
the RTS noise caused by the charge of Q = +1e in the composite gate insulators of
a p-type CNFET may be reduced greatly. 
If we use a combination of the inner insulator with a dielectric constant of 80 and
the outer insulator with a dielectric constant of 3.9 (as for SiO$_2$)
and the inner insulator is much thicker than the outer insultor, the RTS
noise amplitude due to a charge of Q = +1e may be reduced
to no more than 30\% in the turn-on region and 0.1\% in the on-state.

Since a single defect in the gate oxide of a CNFET may cause a large amplitude of RTS noise, it is important to
establish the effective methods to resist radiations; 
on the other hand, CNFETs may also be used for detecting radiations.

\begin{acknowledgments}
We are grateful to S. Heinze for useful discussions.
N-PW acknowledges financial support from The National Natural Science Foundation of China
under Grant No. 61176081 and K.C. Wong Magna Fund in Ningbo University.
\end{acknowledgments}

\newpage

\newpage

\newpage

\newpage

\newpage

\newpage

\newpage

\newpage

\end{document}